\documentstyle[12pt,a4,epsfig]{article}
\newcommand{\beq}{\begin{equation}} \newcommand{\eeq}{\end{equation}} %
\newcommand{\beqar}[1]{\begin{eqnarray}\label{#1}}
\newcommand{\eeqar}{\end{eqnarray}} 

\begin{document}
\vspace*{-2cm}
\begin{flushright}
TPR - 98 - 37
\end{flushright}
\begin{center} {\LARGE \bf
\vspace{2em}
On the Coulomb corrections to the \\
\vspace*{0.2cm}
total cross section of the interaction\\
\vspace*{0.2cm}
of the $\pi^+\pi^-$ atom with ordinary\\
\vspace*{0.2cm}
atoms at high
energy.
}\\[2mm]

\vspace{1cm}
{ \bf D.Yu.~Ivanov$^{a)}$, L.~Szymanowski$^{b)}$ }\\ \vspace{1cm}

\vspace{2em} $^{a)}$ Institute of Mathematics, 630090
Novosibirsk, Russia \\

\vspace{2em} $^{b)}$ Institut
f\"ur Theoretische Physik, Universit\"at Regensburg,
\\ 93040 Regensburg, Germany \\
and \\
Soltan Institute for Nuclear Studies, Ho\.za 69,
00-681 Warsaw, Poland

\end{center}

\vspace{2cm}
\centerline{\bf Abstract:}

The size of $\pi^+\pi^-$ atom in the low lying states
is considerably smaller than the radius of atomic screening.
Due to that we can neglect this screening calculating the
contribution of multi-photon exchanges. We obtain the analytic
formula for Coulomb corrections which works with a
 very good accuracy for the ground state of $\pi^+\pi^-$ atom.

\vspace*{\fill}
\eject
\newpage

1. The proposed accurate  measurement  of the $\pi^+\pi^-$
atom (dimesoatom) lifetime in the experiment DIRAC \cite{dirac}
would give important information about the low energy QCD
dynamics. The process of the breakup (ionization) of the dimesoatom
due to its electromagnetic interaction with the target matter (ordinary
atoms)
is important for this experiment.

The cross section of this process is related to the imaginary
part of the
forward scattering amplitude. At high energy it
is given by the diagrams of Fig.1  with the even number $N=2n$ of photon
exchanges in the t- channel.

\begin{figure}[htb]
\label{fig-1}
\begin{center}
\epsfig{file=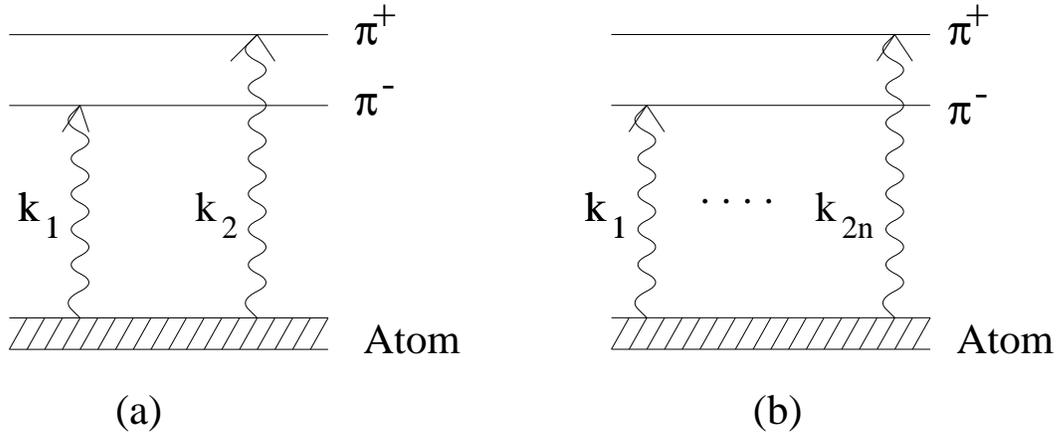,width=14cm}
\caption{Feynman diagrams relevant for the electromagnetic
breakup or relativistic dimesoatom: a) Born  approximation,
b) multiphoton contributions.}
\end{center}
\end{figure}

In the Born approximation this process was considered in
\cite{af96}.
The contributions of the diagrams with
multiphoton exchanges $n>1$, Fig.1(b), is important for an atom having
large $Z$
due to the strong Coulomb field of the nucleus.
The expansion parameter
$\nu=Z\alpha$ is not small in the case of the atom
with large charge $Z$,
therefore the contribution of all diagrams at Fig.1 should be summed
to achieve the exact result.

When our paper was under preparation we learned about
the paper \cite{Gevor} in which a  similar ideas about the
dominance of small transverse distances region
to Coulomb
corrections were proposed. Technically our method is
different from that one used in \cite{Gevor}, it looks for us more
straightforward. Here we
reproduce the result of \cite{Gevor}, see our eqs.(12-15).  Also we calculate
the first correction to this result, see eq.(29).
It gives a possibility to estimate
the accuracy of analytical approach and expand its applicability
region from ground to low lying states of dimesoatom.

Let us divide the total cross section into two parts

\begin{equation}
\sigma=\sigma^{Born}+\sigma^{Coulomb} \ ,
\end{equation}
where $\sigma^{Born}$ is given by the contribution
of the lowest order
two photon exchange amplitude only , see Fig.1(a),
$\sigma^{Coulomb}$
describes the contribution of the diagrams  Fig.1(b)
with $n>1$ and their interference with the Born amplitude, Fig.1(a).

It is known that the sum of type Fig.1
diagrams in high energy kinematics is equivalent to the eikonal
approximation. Recently this process was considered in the
eikonal approach in \cite{AfVo}. In our note we would like to
emphasize the physical picture underlying this process. Namely we discuss
the hierarchy of
transverse distances relevant for various contributions to the cross
section: $\sigma^{Born}$ and $\sigma^{Coulomb}$.
In contrast to
Born part, Coulomb correction receives the main contribution from
the small
distances where the  electromagnetic field of atom is determined by the
Coulomb field of nuclei. Therefore $\sigma^{Coulomb}$ can be calculated
with high precision since the details of atomic screening in this case  are
of a
small importance.
Our consideration will be very similar to that in
\cite{IvMe}, where the closely related problem, lepton pair
photoproduction in a strong Coulomb field, was considered.
We derive the
analytical formula which gives with good accuracy
$\sigma^{Coulomb}$ for the ground state and
describes qualitatively  low lying dimesoatom states.

Electromagnetic field of atom consists of the field of nucleus
and the field of electron shell. The nucleus field is screened by the
electron shell at large distances
\begin{equation}
r\sim r_A=\frac{1}{m_e\alpha Z^{1/3}} \ .
\end{equation}
At very small distances
\begin{equation}
r\leq r_N=\frac{1}{\Lambda} \ , \Lambda \approx 30 \mbox{ MeV } \ ,
\end{equation}
electromagnetic field depends on the distribution of electric
charge inside nucleus. There is very broad region
\begin{equation}
r_N \; < \; r \; < \; r_A
\end{equation}
where electromagnetic field of atom coincides with the Coulomb
law $\frac{Z\alpha}{r}$.

The other important for our problem dimensional
parameter is the distance
between $\pi^+$ and $\pi^-$ in dimesoatom, which is
in a good approximation positronium--like weakly bound state,
\begin{equation}
r_{2\pi}\sim \frac{2}{m_\pi \alpha} \ .
\end{equation}
It is important to note that this parameter lies in between
of the parameters describing atomic and nuclear screening
\begin{equation}
r_N<< r_{2\pi} << r_A \ .
\end{equation}

Let us discuss now the relevant transverse momenta $< k_i >$ in the
integrals
describing the diagrams at Fig.1, and therefore, the important
for our process transverse distances $r\sim 1/<k_i>$ between nucleus and
high energy dimesoatom.
These $<k_i>$ are different for the various contributions to the cross
section. Calculating Born part $\sigma^{Born}$ we meet
the logarithmic--type integral collecting from the region
\begin{equation}
\frac{1}{r_A}\;\; \leq \;\;<k_{1,2}> \;\;\leq \;\;\frac{1}{r_{2\pi}} \ .
\end{equation}
Therefore  large distances, $r\sim r_A$, where the Coulomb field
of nucleus is screened by the electron shell give sizeable contribution to
the
$\sigma^{Born}$.
On the other hand \\ $<k_i>\sim \frac{1}{r_{2\pi}}$ in the case
of Coulomb contribution to the cross section and hence
the typical transverse
distances for $\sigma^{Coulomb}$  are of order of the
dimesoatom size $r_{2\pi}$. Since there is a large gap between
this size and $r_A$ and $r_N$
in calculation of $\sigma^{Coulomb}$
we can safely neglect the nuclear screening and with a good
approximation the atomic screening.
The accuracy of this approximation will be discussed later.

2. We start with eq.(1) of \cite{AfVo}, see also similar equations
(48),(52) in \cite{IvMe},

\begin{equation}
\label{e1}
\sigma_{nlm} = 2 \, \mbox{Re} \int d^2 b\: d^3 r
|\psi_{nlm}(\vec{r})|^2 \;\left[1-\exp{\left(i\chi(\vec{b}-\vec{s}/2) -
i\chi(\vec{b}+\vec{s}/2)\right)}\right] \,.
\end{equation}
Here $\vec s= \vec r_\perp$ is the projection of the vector $\vec r$ on the
plane perpendicular to
 the collision axis,
the impact parameter  of dimesoatom is  $\vec b$,
$\psi_{nlm}(\vec{r})$
is the wave function of $\pi^+\pi^-$ atom in the state with principal, orbital
and magnetic quantum numbers $n$, $l$ and $m$ respectively. The phase shift
$\chi(\vec b)$ is expressed via  potential of the target
atom:
\begin{equation}
\label{e2}
\chi(\vec b)= \int\limits_{-\infty}^{\infty}
U(\sqrt{b^2+z^2}~)\:dz\,.
\end{equation}
The difference between the phase shifts of $\pi^+$ and
$\pi^-$ can be easily calculated in the case of Coulomb potential.
Therefore
\begin{equation}
\label{e11}
\sigma_{nlm} = 2 \, \mbox{Re} \int d^2 b\: d^3 r
|\psi_{nlm}(\vec{r})|^2 \;\left[1-
\left(\frac{(\vec{b}-\vec{s}/2)^2}{(\vec{b}+\vec{s}/2)^2}
\right)^{i\nu} \right] \,.
\end{equation}
This integral becomes convergent after the subtraction of the
Born contribution which is divergent in the  case of
unscreened Coulomb potential.
\begin{equation}
\label{e111}
\sigma_{nlm}^{Coulomb} = 2 \, \mbox{Re} \int d^2 b\: d^3 r
|\psi_{nlm}(\vec{r})|^2 \;\left[1-
\left(\frac{(\vec{b}-\vec{s}/2)^2}{(\vec{b}+\vec{s}/2)^2}
\right)^{i\nu} - \frac{\nu^2}{2}
\ln^2 (\frac{(\vec{b}-\vec{s}/2)^2}{(\vec{b}+\vec{s}/2)^2} )\right] \,.
\end{equation}
After the substitution $\vec{b}\to s(\vec{R}-\vec{n})$, where $\vec{n}=
\frac{\displaystyle \vec{s}}{\displaystyle s}$,
the integral in (\ref{e111}) factorizes, see also \cite{IvMe}.
\begin{equation}
\label{e12}
\sigma_{nlm}^{Coulomb} =<s^2>I_{\nu} \ ,
\end{equation}
where
\begin{equation}
\label{e13}
I_\nu=\int d^2 \vec{R}
\left\{
2
-
\left(
\frac{R^2}{(\vec{R}-\vec{n})^2}
\right)^{i\nu}
-
\left(
\frac{R^2}{(\vec{R}-\vec{n})^2}
\right)^{-i\nu}
-\nu^2\ln^2
\left(
\frac{R^2}{(\vec{R}-\vec{n})^2}
\right)
\right\}
\end{equation}
and
\begin{equation}
\label{e14}
<s^2>
=\int d^3r r^2\sin^2\Theta |\psi_{nlm}(\vec{r})|^2=
<r^2>_{(n,l)}<\sin^2\Theta >_{(l,m)}
\end{equation}
The integral $I_\nu$ was calculated in \cite{IvMe}
\begin{equation}
\label{e15}
I_\nu=-4\pi\nu^2f(\nu) \ , \
f(\nu)=\frac{1}{2}
[\Psi (1-i\nu)+ \Psi (1+i\nu) -2\Psi (1)] \ , \
\end{equation}
where $\Psi(z)=d(\ln \Gamma(z))/dz $.
Note that the dependence of the $\sigma^{Coulomb}$
on $Z$  factorizes from the variables describing
the state of $\pi^+\pi^-$ atom. It is given by the
universal function $f(\nu)$.

$<r^2>_{(n,l)}$ for the positronium--like states
is given by \cite{landau}
\begin{equation}
<r^2>_{(n,l)}=\left( \frac{2}{m_\pi\alpha}\right)^2 \frac{n^2}{2}
[5n^2+1-3l(l+1)]
\end{equation}
We will consider for simplicity
the cross section averaged over the magnet
quantum number:
\begin{equation}
\label{e10}
\sigma_{nl} =\frac{1}{2l+1} \sum_{m} \sigma_{nlm} \, .
\end{equation}
In this case
\begin{equation}
<\sin^2 \Theta >=2/3 \ .
\end{equation}
Taking into account all factors   we find the result
\begin{eqnarray}
&&\sigma_{nl} = \sigma_{nl}^{Born} + \sigma_{nl}^{Coulomb} \nonumber \\
&& \sigma_{nl}^{Coulomb} =
-\frac{16\pi\nu^2 f(\nu)}{m_\pi^2 \alpha^2}
\frac{n^2}{3}[5n^2+1-3l(l+1)] \ .
\label{result}
\end{eqnarray}

3. Note that $\sigma^{Coulomb}$ is proportional to
$r^2_{2\pi}$, the mean square of the distance between $\pi^+$ and $\pi^-$
in the dimesoatom.  $r^2_{2\pi}$   grows rapidly, $\sim n^4$,
with increasing $n$
for weakly bounded dimesoatom.
At $n\sim 4$ the distance between $\pi^+\pi^-$ becomes of  the
order of the radius of atomic screening $r_A$ for atom with large $Z$.
Therefore our approach based on the large difference between $r_{2\pi}$
and $r_A$
can not be applied to the highly excited states of dimesoatom.

The  first correction 
related to  appearance  of the atomic
screening can also be calculated analytically. In order to
evaluate it let us replace the denominator of the photon
propagator by
\begin{equation} \label{sub} \frac{1}{{\bf k}^2} \to
\frac{1}{{\bf k}^2 + \mu^2}\;,
\end{equation}
where $\mu$ is the
inverse of the radius of atomic screening $r_A$.  This
replacement leads to the analog of the formula (\ref{e111})
having now the form

\begin{eqnarray}
\label{e311}
\sigma_{nlm}^{Coul.+Atom. Scr.} &=& 2 \, \mbox{Re} \int d^2 b\: d^3 r
|\psi_{nlm}(\vec{r})|^2 \;\left[1-
e^{2i\nu [K_0(\mu|\vec{b} - \frac{\vec{s}}{2}|) -
K_0(\mu|\vec{b} + \frac{\vec{s}}{2}|)]}  \right.\nonumber \\
&-&\left. 2\nu^2 [K_0(\mu|\vec{b} - \frac{\vec{s}}{2}|) -
K_0(\mu|\vec{b} + \frac{\vec{s}}{2}|)]^2 \right]\;,
\end{eqnarray}
where $K_0(z)$ is the modified Bessel function.

The  first  correction
due to the atomic screening can be obtained from
eq.(\ref{e311}) by taking into account that for small values of $\mu$
\begin{eqnarray}
\label{K}
&&K_0(\mu|\vec{b} - \frac{\vec{s}}{2}|) -
K_0(\mu|\vec{b} + \frac{\vec{s}}{2}|)=
 \ln \frac{|\vec{b} + \frac{\vec{s}}{2}|}{|\vec{b}
- \frac{\vec{s}}{2}|}
+ \frac{\mu^2}{4}\left[ (|\vec{b} - \frac{\vec{s}}{2}|^2 -
|\vec{b} + \frac{\vec{s}}{2}|^2 )\Psi(2) \right. \nonumber \\
&&+ \left.
|\vec{b} + \frac{\vec{s}}{2}|^2 \ln \frac{\mu |\vec{b} +
\frac{\vec{s}}{2}|}{2}
-|\vec{b} - \frac{\vec{s}}{2}|^2 \ln \frac{\mu |\vec{b} -
\frac{\vec{s}}{2}|}{2}
\right] \;,
\end{eqnarray}
and keeping terms proportional to $\mu^2$.
In this way we obtain
\begin{eqnarray}
\label{corr}
&&\Delta \sigma_{nlm}^{Atom. Scr.} = -2\mu^2 \, \mbox{Re} \int d^2 b\: d^3 r
|\psi_{nlm}(\vec{r})|^2 \;\left[\frac{i\nu}{2}
e^{2i\nu \ln \frac{|\vec{b} + \frac{\vec{s}}{2}|}{|\vec{b}
- \frac{\vec{s}}{2}|}} + \frac{\nu^2}{2} \ln \frac{|\vec{b}
+ \frac{\vec{s}}{2}|^2}{|\vec{b} - \frac{\vec{s}}{2}|^2}\right]\cdot \nonumber \\
&&\left[( |\vec{b} - \frac{\vec{s}}{2}|^2 -
|\vec{b} + \frac{\vec{s}}{2}|^2 )\Psi(2) +
|\vec{b} + \frac{\vec{s}}{2}|^2 \ln \frac{\mu |\vec{b} +
\frac{\vec{s}}{2}|}{2}
-|\vec{b} - \frac{\vec{s}}{2}|^2 \ln \frac{\mu |\vec{b} -
\frac{\vec{s}}{2}|}{2}
\right] \:.
\end{eqnarray}
The dominant contribution to this expression comes from the region of values
of $\vec{b}$ being much lager than the size of the dimesoatom described by
$\vec{s} = \vec{r}_\perp$. In this limit we obtain
\begin{equation}
\label{dcorr}
\Delta \sigma_{nlm}^{Atom. Scr.} \approx  \frac{8}{3}\mu^2  \nu^4 \, \int \: d^3
r\: \int\limits_r^{r_A} \:d^2 b |\psi_{nlm}(\vec{r})|^2 \;
\frac{(\vec{b}\vec{s})^4}{(\vec{b}^2)^3}\left[ \Psi(2) - \frac{1}{2}\ln
\frac{\mu^2 \vec{b}^2}{4} - \frac{1}{2}\right] \:,
\end{equation}
and consequently  with the logarithmic accuracy we can write
\begin{equation}
\label{dcorrection}
\Delta \sigma_{nlm}^{Atom. Scr.} \approx  \pi\mu^2  \nu^4 \, \int \:
d^3 r\:|\psi_{nlm}(\vec{r})|^2 \; r^4 \sin^4 \Theta \ln^2 (\mu r)\;.
\end{equation}

This result can be generalized to the case in which there appear a sum of
several potentials, i.e. when instead of eq.(\ref{sub}) we perform the
substitution
\begin{equation}
\frac{1}{{\bf k}^2} \to \sum_i^N
\frac{c_i}{{\bf k}^2 + \mu^2_i}\:,\;\;\;\sum_i^N c_i = 1\:.
\end{equation}
This generalization includes, in particular,
the  Moli\'ere \cite{Moliere} parametrization of the
Thomas-Fermi potential. In this case the
eq.(\ref{dcorrection}) generalizes to the formula
\begin{equation}
\label{sumdcorr}
\Delta \sigma_{nlm}^{Atom. Scr.} \approx  \pi \nu^4 \int \: d^3
r\:|\psi_{nlm}(\vec{r})|^2 \; r^4 \sin^4 \Theta \sum_i^N \frac{1}{4}c_i \mu^2_i
\ln^2(\mu_i^2 r^2) \:.
\end{equation}
Since the derivation of eq.(\ref{dcorrection}) was performed
with the logarithmic accuracy
we are free to put in the argument of logarithmic function in
eq.(\ref{sumdcorr}) some average value $\bar{\mu}^2$ of the square
of masses $\mu_i$.  We choose as $\bar{\mu}^2$ the quantity
\begin{equation}
\bar{\mu}^2 = \sum_i^N c_i \mu_i^2\:.
\end{equation}
In this way we arrive to the final form of correction related to the atomic
screening averaged over the magnet quantum number $m$ (see eq.(\ref{e10}))
\begin{eqnarray}
\label{finalscr}
&&\Delta \sigma_{nl}^{Atom. Scr.} \approx  \frac{1}{4} \pi \nu^4  (\sum_i^N
c_i \mu^2_i )\: \frac{1}{2l+1}\sum_{m=-l}^{l}\;\int \: d^3
r\:|\psi_{nlm}(\vec{r})|^2 \; r^4  \ln^2 (\bar{\mu}^2 r^2)  \sin^4 \Theta
\nonumber \\
&& \approx \frac{2}{15} \pi \nu^4  (\sum_i^N c_i \mu^2_i )\:<r^4>_{n,l}
\ln^2 (\bar{\mu}^2 <r^2>_{n,l}) \:
\end{eqnarray}
The cross section under consideration is given by the sum of terms
\begin{equation}
\sigma_{nl} = \sigma_{nl}^{Born} + \sigma_{nl}^{Coulomb} + \Delta
\sigma_{nl}^{Atom. Scr.}\:,
\end{equation}
where $\sigma_{nl}^{Coulomb}$ is given by eqs. (\ref{result}) and
(\ref{e15}),  and
$\Delta\sigma_{nl}^{Atom. Scr.}$ is given by eq.(\ref{finalscr}).

Let us now compare the results of numerical calculations for
$(\sigma_{nl}^{Born} - \sigma_{nl})/\sigma_{nl}^{Born}$ in the case of
Tantalum ($Z=73$) presented on Fig.2
of Ref. \cite{AfVo} with the analytic results obtained above. The values of
$\sigma_{nl}^{Born}$ for different values of $n$ and $l=0$ we take from the first
column of the Table in Ref. \cite{af96}. The values of constants $c_i$ and
 masses $\mu_i$ are  equal to those in Ref. \cite{AfVo} and read

\begin{eqnarray}
&&c_1 = 0.35 \;\;\;\;c_2 = 0.55 \;\;\;\; c_3=0.1 \nonumber \\
&&\mu_1 =0.3\mu_0 \;\;\;\; \mu_2 =1.2 \mu_0 \;\;\;\; \mu_3=6\mu_0
\\
&&\mbox{where} \;\;\; \mu_0 = \frac{m_e \alpha Z^{1/3}}{0.885}\;.
\end{eqnarray}
Table 1 shows the result of this  comparison.

\begin{table}
\begin{center}
\caption{ Comparison  of the numerical results of
Ref.\cite{AfVo} with the analytic results.} \vspace*{0.2in}
\begin{tabular}{|c|c|c|c|}
\hline
           & Fig.2 of Ref.\cite{AfVo} & eq.(19) & eq.(30)\\ \hline
$n=1, l=0$ & 0.082 & 0.0851 & 0.08212 \\ \hline
$n=2, l=0$ & 0.109 & 0.1321 & 0.1265 \\ \hline
$n=3, l=0$ & 0.128 & 0.1962 & 0.1961 \\ \hline
$n=4, l=0$ & 0.138 & 0.2846 & 0.1812 \\
\hline
\end{tabular}
\end{center}
\end{table}

We see that for the ground state of dimesoatom
the first atomic screening
correction reduces practically to zero the
difference between result of the numerical
calculation of Ref.\cite{AfVo} and our analytic result (19). For the exited
states this correction leads to  decreasing of
this difference but the  precision is
low. Consequently,
 for those states the analytic approach based on eqs. (19) or (30) works only qualitatively.

4. We want still to add  two additional remarks. First, let us note
that
the nuclear screening can be really safely neglected
since the corresponding relative accuracy is very high
\begin{equation}
\sim \left( \frac{r_N}{r_{2\pi}}\right)^2 < 10^{-4} \ .
\end{equation}

Secondly,  due to the fact that the values
of muon and pion masses are close to each other,
our qualitative consideration and its  consequence given by
eqs. (\ref{result}),(\ref{finalscr})
are directly applicable to the process of the breakup of
$\mu^+\mu^-$ atom.

\vspace*{1cm}

{\Large \bf  Acknowledgments:}

\centerline{}
D.Yu.I. would like to acknowledge the hospitality at Soltan Institute for
Nuclear Studies in
Warsaw where this work started. His work is also supported by
Russian Foundation for Basic Research , grant N 96-02-19114.

L.Sz. would like to acknowledge the support from German-Polish agreement on
scientific and technological cooperation N-115-95.

\end{document}